\documentstyle[12pt]{article}

\parskip=\medskipamount
\begin{document}

\begin{titlepage}

\begin{center}
{ \large \bf Centauro- and anti-Centauro-type events }\\
\vspace{2.cm}
{ \normalsize \bf M. Martinis, \footnote { e-mail address:
martinis@thphys.irb.hr}
 V. Mikuta-Martinis, \footnote {e-mail address: mikutama@thphys.irb.hr}
A. \v Svarc, and J. \v Crnugelj \footnote
{e-mail address:crnugelj@thphys.irb.hr }} \\
\vspace{0.5cm}
Department of Theoretical Physics, \\
Rudjer Bo\v skovi\' c Institute, P.O.B. 1016, \\
41001 Zagreb,CROATIA \\
\vspace{2cm}
{ \large \bf Abstract}
\end{center}
\vspace{0.5cm}
\baselineskip=24pt

Assuming that leading particles in high-energy
hadronic and nuclear collisions
become sources of a classical pion field, we
show that the direct production of pions favors Centauro
( mainly charged ) events and that the production of pions
through the $ \rho$-type channel favors anti-Centauro
( mainly neutral ) events. We also observe a strong
negative neutral-charged correlation in both cases.
\end{titlepage}

\renewcommand{\thesection}{\Roman{section}}
\baselineskip=24pt
\setcounter{page}{2}
\section{Introduction}
Recently, there have been several interesting theoretical
speculations  [1--6] that localized regions of misaligned vacuum
might occur in ultrahigh-energy hadronic and heavy-ion
collisions. These regions become
coherent sources of a classical pion field.
The models in [1--6] predict large isospin fluctuations on
the event - to - event basis, in particular the
fluctuation in the neutral - to - charged ratio,
thus offering a possible
explanation of Centauro cosmic-ray experiments [7].
Although the actual dynamical mechanism of the
production of a classical pion field in the course
of a high-energy collision is not known, there exist a
number of calculations in which the coherent
production of pions is considered to be a dominant mechanism
[8,9]. These early models also predict strong negative
correlations between the number of neutral and
charged pions. In fact, the exact conservation of isospin
in a pion uncorrelated jet model is known [10,11] to
give the same pattern of charged/neutral fluctuations
as observed in Centauro events. This strong negative
neutral-charged correlation is believed to be a
general property of the direct pion emission in
which the cluster formation is not taken into account
[12,13,14].

The same conclusion is drawn when a proper
multipion symmetrization combined with isospin is
considered [15].

In this paper we consider the leading-particle effect as a
possible source of a classical pion field. Pions are
produced from a definite isospin state of the
incoming leading-particle system either directly
or through the cluster emission mechanism.

Coherent emitted clusters decay subsequently into
pions outside the region of interaction. We discuss
the behavior of the probability distribution of the
neutral - to - charged ratio and the corresponding
two-pion correlation functions $C_{ab}$ when the total
number of produced pions is finite but large.

\newpage
\section{Coherent production}

Many results on hadronic and nuclear collisions can be understood
in terms of a simple picture that the outgoing particles have three
origins: beam fragmentation, target fragmentation, and
central production. At high energies most of the pions are
produced in the central region. To isolate the central production,
we adopt high-energy longitudinally dominated kinematics, with
two leading particles retaining a large fraction of their incident
momenta.

With a set of independent variables $s, \; \{
\vec{q}_{iT},y_{i} \} \equiv q_{i}, \; \; i = 1,2, \ldots n,$
the n-pion contribution to the $s$-channel unitarity becomes an
integral over the relative impact parameter  $b$ of the two
incident leading particles:
\begin{equation}
F_{n}(s) = \frac{1}{4s} \int d^{2}b \prod_{i=1}^{n}dq_{i}
\mid T_{n}(s, \vec{b};1 \ldots n) \mid^{2},
\end{equation}
where $ dq = d^{2}q_{T}dy/2(2 \pi)^{3}.$ The normalization is
such that
\begin{eqnarray}
F_{n}(s) & = & s \sigma_{n}(s), \nonumber \\
\sigma_{inel}(s) & = & \sum_{n=1}^{ \infty} \sigma_{n}(s).
\end{eqnarray}

The leading-particle effect is crucial for an approximate
treatment of the multiparticle $s$-channel unitarity integral,
Eq.(1),  which enables us to consider the colliding particles
as a classical source of pions in the following way.
At high energies the matter
distributions of the two-leading-particle
system are Lorentz-contracted disks
in the center - of - mass system. A highly excited localized system
occurs, just after the collision and then relaxes through the coherent
emission of pions
[16]. According to [17], this type of coherent emission of
pions should saturate close to the threshold since,
once it starts, the resulting pion emission, in the
absence of a " resonance cavity ", prevents further
buildup of pion fields. The basic equation for the
pion field is
\begin{equation}
( \Box + \mu^{2}) \vec{ \pi}( s,\vec{b}; x) = \vec{j}(s,
\vec{b};x),
\end{equation}
where $ \vec{j}$ is a classical source. If the source
$ \vec{j}(s, \vec{b},x)$ is static, i.e., independent of time,
no pion can be radiated. However, in the dynamical case
of hadronic or nuclear collisions, $ \vec{j}$ acquires time
dependence and radiates pions. The reference to
the initial leading-particle system is contained in the variables
$ \vec{b}$ and $s$.
The standard solution of Eq.(3) is given in terms of in- and
out-fields that are connected by the unitary
$S$ matrix $ \hat{S}( \vec{b},s)$ as follows:
\begin{equation}
\vec{ \pi}_{out} = \hat{S}^{ \dagger} \vec{ \pi}_{in}
\hat{S} = \vec{ \pi}_{in} + \vec{ \pi}_{classical}.
\end{equation}

The $S$ matrix following from such a classical source
is still an operator in the space of pions. On the
other hand, inclusion of isospin requires $ \hat{S}(s, \vec{b})$
to be also a matrix in the isospace of the leading particles.

If the isospin of the system of two incoming ( outgoing )
leading particles is $II_{3}$ and $( I'I'_{3})$, respectively,
then the initial-state
vector for the pion field is $ \hat{S}( s, \vec{b}) \mid II_{3} \rangle$,
where  $ \mid II_{3} \rangle$ is a vacuum state with no pions
but with two leading particles in the isostate characterized by $II_{3}.$
Then the $n$-pion production amplitude is
\begin{equation}
iT_{n}(s, \vec{b}; q_{1} \ldots q_{n}) = 2s \langle
I'I'_{3}; q_{1} \ldots q_{n} \mid \hat{S}(s,
\vec{b}) \mid II_{3} \rangle.
\end{equation}
The unnormalized probability distribution of producing
$n_{+} \pi^{+}, n \_ \pi^{ -},$ and $ n_{0} \pi^{0}$ pions
is defined as
\begin{eqnarray}
W(n_{+}n \_ n_{0},I'I'_{3},II_{3}) & = & \int d^{2}bdq_{1}dq_{2}
\ldots dq_{n} \mid \langle I'I'_{3}n_{+}n \_ n_{0} \mid
\hat{S}(s, \vec{b}) \mid II_{3} \rangle \mid^{2}, \\
\mbox{where} \hspace{1.5cm} n & = & n_{+} + n \_ + n_{0}. \nonumber
\end{eqnarray}
The coherent production of pions from a
classical source [ Eq. (3) ] is described by the following $S$ matrix:
\begin{equation}
\hat{S}(s, \vec{b}) = \int d^{2} \vec{e} \mid \vec{e}
\rangle D( \vec{J}; s, \vec{b}) \langle \vec{e} \mid,
\end{equation}
where $ \mid \vec{e} \, \rangle $ represents the isospin-state
vector of the two-leading-particle system. It has the property
that
\begin{eqnarray}
\langle \vec{e} \mid \vec{e} \, ' \rangle  =  \delta^{(2)}(
\vec{e} - \vec{e} \, '), & & \nonumber \\
\int \mid \vec{e} \, \rangle d^{2} \vec{e}
\langle \vec{e} \mid \; = \; 1, \hspace{.5cm} & & \\
\langle \vec{e} \mid II_{3} \rangle = Y_{I{I_{3}}}(
\vec{e}), \hspace{.5cm} & &
\end{eqnarray}
where $ Y_{I{I_{3}}}( \vec{e})$ is the usual spherical harmonic.
The quantity $D( \vec{J};s, \vec{b})$ is the unitary coherent-state
displacement operator defined as
\begin{equation}
D( \vec{J};s, \vec{b}) = exp[ \int dq \vec{J}(s, \vec{b};q)
\vec{a}^{ \dagger}(q) -H.c.],
\end{equation}
where $ \vec{a}^{ \dagger}(q)$ is the creation operator of
a physical pion and
\begin{equation}
\vec{J}(s, \vec{b};q) = \int d^{4}x e^{ iqx}
\vec{j}(s, \vec{b};x)
\end{equation}
is the Fourier transform of a classical pion source.
Note that each charged state of the pion field has a  source
that varies arbitrarily in space and time. However, if
we assume pions to be identical, then the isospin of
all pions, regardless of their momenta, is coupled to form the
total isospin. This is obtained by considering a coherent
production with $ \vec{J}(s, \vec{b}; q)$ of the form
\begin{equation}
\vec{J}(s, \vec{b};q) = J(s, \vec{b};q) \vec{e},
\end{equation}
where $ \vec{e}$ is a fixed unit vector in isospace. At this
point the conservation of isospin becomes a global property
of the system, restricted only by the relation
\begin{eqnarray*}
\vec{I} = \vec{I}' + \vec{I}_{ \pi}, \nonumber
\end{eqnarray*}
where $ \vec{I}_{ \pi}$ denotes the isospin of the emitted pion
cloud. In this model, the pions in the cloud are uncorrelated
and described by the same wave function in momentum space for
each pion.

Assuming further that the total number of emitted pions
is finite, but large and that all $(I', I_{3}')$ are produced with
equal probability, we can sum over all possible isospin states of
the outgoing leading particles to obtain \\
\begin{equation}
P_{I{I_{3}}}(n_{+}n \_ n_{0} \mid n ) = \frac{
\sum_{I'I'_{3}}W(n_{+}n\_ n_{0},I'I'_{3};II_{3})}{
\sum_{n_{+}+n \_ +n_{0}=n} \sum_{I'I'_{3}} W(n_{+} n \_
n_{0}, I'I'_{3};I I_{3} )}.
\end{equation} \\
This is  our basic relation for calculating various pion
multiplicity distributions, pion multiplicities, and pion
correlations between definite charge combinations.
In general, the probability $W(n_{+}n_{ \_}n_{0},I'I_{3}';II_{3})$
depends on $(I'I_{3}')$ dynamically. The final-leading-particle
production mechanism usually tends to favor the $(I',I_{3}') \approx
(I,I_{3})$ case, for example, if the final-leading particles are
nucleons or isobars. However, if the leading particles are
colliding nuclei, it is resonable to assume almost equal
probability for various $(I',I_{3}')$ owing to the large number
of various leading isobars in the final state.

Note that, in general, the $n-$pion cloud contains components of
all isospins: $I_{ \pi} = 0,1, \dots, n.$
The case when both the isospin state
of the incoming- and that of the outgoing-leading-particle
systems are fixed has been treated in [12].

\section{Grey-disk dynamics}

In order to obtain some detailed results for multiplicity
distributions and correlations, we have to choose an explicit
form for the source function $J(s, \vec{b};q).$

The results on the isospin structure are most easily
analyzed in the grey-disk model in which
\begin{equation}
\int dq \mid J(s, \vec{b};q) \mid^{2} = \Delta \theta(b_{o}-b).
\end{equation}
Here $ \Delta$ and $b_{0}$ are in general energy-dependent
parameters and $ \theta$ is a step function.
The parameter $ \Delta$ is expected to
grow linearly with $ \ln s$ in the central region. Such behavior is also
expected from a multiperipheral model.

For $ I_{3} = I,$ it is a straightforward algebra to calculate
the probability of creating $n$ pions of
which $n_{0}$ are neutral pions:
\begin{eqnarray}
P_{I}^{( \pi)}( n_{0} \mid n ) & = & \sum_{{n_{+}}+n \_
= n-n_{0}}P_{II}(n_{+}n \_ n_{0} \mid n) \\
& = & \left( \begin{array}{c}
n \\ n_{0} \end{array} \right) \frac{B(n_{0}+ \frac{1}{2},
n-n_{0}+I+1)}{B( \frac{1}{2},I+1)}.
\end{eqnarray}
Here $B(x,y) = \frac{ \textstyle \Gamma(x) \Gamma(y)}{
\textstyle \Gamma(x+y)}$
is the Euler beta function. Note that $P_{I}^{( \pi)}(
n_{0} \mid n)$ differs considerably from the binomial
distribution given here by $3^{-n}2^{n-n_{0}}( \stackrel{
\textstyle n}{n_{0}}).$
In Fig. 1 we show the  behavior of $P_{I}^{( \pi)}$
for $n=100$ and for different isospin $(I=0,1, \ldots$ )
of the initial-leading-particle system.
Figure 1 also shows a comparison  with the corresponding binomial distribution.
It is clear that our direct-pion-emission model predicts
many more events with a small number of neutral pions
(Centauro events), in particular if the isospin of the
initial-leading-particle system is large.

\newpage

\section{Neutral - to -  charged ratio}

{ \bf a) Centauro events}

Let us define $R= \frac{ \textstyle n_{0}}{
\textstyle n}$, the fraction of neutral
pions in an event with $n$ pions. Then it is easy to see
that in the limit $ n \rightarrow \infty$, with $R$ fixed,
the probability distribution $nP_{I}^{( \pi)}(n_{0} \mid n)$
scales to the limiting behavior:
\begin{equation}
nP_{I}^{( \pi)}(n_{0} \mid n) \rightarrow P_{I}^{( \pi)}(R) =
\frac{(1 - R)^{I}}{B( \frac{1}{2},I+1) \sqrt{R}}.
\end{equation}
This limiting probability distribution is different from
the usual Gaussian random distribution for
which one expects $P_{I}^{( \pi)}(R)$ to be peaked at
$R = \frac{1}{3}$ as $n \rightarrow \infty.$

Instead, we find peaking at $R=0$ although
\begin{eqnarray}
\begin{array}{rl}
\langle n_{0} \rangle & = \; n - \langle n_{ch} \rangle \\ \\
& = \; \frac{ \textstyle 1}{ \textstyle 2I+3}n \\ \\
\mbox{and} \hspace{2cm} \langle n_{0} \rangle & =
\frac{1}{2} \langle n_{ch} \rangle = \frac{1}{3}n, \hspace{1cm}
\mbox{for $I$ = 0.}
\end{array}
\end{eqnarray}

The distribution (17) for $I=0$ is similar to the distribution that has been
derived and advocated in the disoriented chiral condensate model [2].
Although the final distributions of pions look the same, the underlying
mechanism is different. In the disoriented chiral model, the classical
pion field satisfies a nonlinear equation of motion that can be reduced
to the equation with a time-dependent effective mass:
\begin{eqnarray*}
( \Box + \mu_{eff}^{2}(t)) \vec{ \pi}(x) & = & 0,
\end{eqnarray*}
in contrast to our Eq.(3).

In addition, there is also a strong isospin correlation
between various pion-pair channels:
\begin{eqnarray}
C_{00}^{(I)} & = & 4 \frac{I +1}{2I + 5}, \nonumber \\
C_{cc}^{(I)} & = & \frac{1}{(I +1)(2I + 5)}, \\
C_{0c}^{(I)} & = & - \frac{2}{2I + 5}, \nonumber
\end{eqnarray}
\begin{eqnarray*}
\mbox{where} \hspace{1.5cm} C_{ab}^{(I)} =
\frac{ \langle n_{a}n_{b} \rangle}{
\langle n_{a} \rangle \langle n_{b} \rangle } -1.
\end{eqnarray*}
Note the strong negative neutral-charged correlation,
which is a general property of the direct pion
production. Experimentally, this correlation is
positive at least at high energies.

{ \bf b) Anti-Centauro events}

Let us now assume that pions are produced through the coherent
production of clusters that decay into two or more pions.
The more pions in a cluster, the larger the correlation effect
expected. Here we restrict our discussion to
isovector clusters of the $ \rho $ type. The pion
distribution and correlations follow from (15) by
observing that
\begin{eqnarray}
n_{ \rho +} & = & \frac{1}{2}n - n \_, \nonumber \\
n_{ \rho \_ } & = & \frac{1}{2}n - n_{+}, \\
n_{ \rho 0 } & = & \frac{1}{2}n - n_{0}. \nonumber
\end{eqnarray}
We find that
\begin{equation}
P_{I}^{( \rho)}(n_{0} \mid n) = P_{I}^{( \pi)}(
\frac{1}{2}n - n_{0} \mid \frac{1}{2}n),
\end{equation}
where $ n = 2,4, \ldots$ is even.

In the limit $ n \rightarrow \infty$, with $R = n_{0}/n$
fixed, we find the following behavior: \\
\begin{equation}
nP_{I}^{( \rho)}(n_{0} \mid n) \rightarrow
P_{I}^{( \rho)}(R) = \frac{2(2R)^{I}}{B(
\frac{1}{2},I+1) \sqrt{1-2R}}.
\end{equation} \\
This distribution is sharply peaked at $R = \frac{1}{2}$,
i.e., when $n_{0} = \frac{1}{2}n.$ According to (22),
there is a substantial probability of events
only with neutral pions ( anti-Centauro events ).
The neutral-charged correlations are found to be
less negative than in the case of direct pion
production:
\begin{equation}
C_{0c}^{(I)} = \frac{-1}{(I+2)(2I+5)}.
\end{equation}

In Fig. 2 we show the behavior of (21) for
$n=100,$ and $I = 0,1,2,3.$

\section{Conclusion}

Assuming that leading particles in high-energy hadronic and
nuclear collisions become a
source of a classical pion field, we have shown
that the coherent production of pions
favors Centauro-type events, whereas the coherent
production of $ \rho$ - type clusters will favor
anti-Centauro-type events. In the anti-Centauro case, $n/n_{0}$
peaks near 1. The pions in the events of $n \sim n_{0}$
contain very large total isospins of $I_{ \pi} \sim n.$
Therefore the final leading particle must also include isospins
of $I' \approx I_{ \pi},$ to match the initial isospin $I$ if it is
small. That is, existance of the peak near $n_{0}/n = 1$ in
$P(n_{0} \mid n)$ requires abundance of very high $I'$ states
which is possible in heavy ion collisions.
The same argument can be made for the Centauro case.
We also predict a
strong negative neutral-charged correlation.
It is clear that one should consider both
$ \pi, \; \rho,$ and other multipion
cluster production to obtain a realistic
picture of the classical pion production.

In addition, the chosen form of the source function
of the classical pion field, Eq. (14), is too simple.
One should probably try to find a more adequate space-time
structure leading to the pion condensate, for example [17].
Another possibility is to relate the pion-source function
to the single-particle inclusive
distribution. We hope to address this question
in the near future.
\vspace{1cm}

{ \large \bf Acknowledgment }

This work was supported by the Ministry of Science of
Croatia under Contract No. 1 - 03 - 212.

\newpage

{\bf Figure captions :}

Fig. 1. Multiplicity distributions $P_{I}^{( \pi)}
(n_{0} \mid n)$ of neutral pions for $n=100$ when
the total isospin of
the ingoing-leading-particle system
is $I = 0,1,2,3.$ The dashed line represents the
corresponding binomial distribution with the isospin
invariance neglected.

Fig. 2. Same as Fig. 1, except that pions are
assumed to be
emitted through the isovector $ \rho$ - type
channels.

\end{document}